# Laser-induced THz magnetism of antiferromagnetic $CoF_2$

F. Formisano[1*], R. M. Dubrovin[2], R. V. Pisarev[2], A. M. Kalashnikova[2] and A. V. Kimel[1]

[1]Institute for Molecules and Materials, Radboud University, 6571AJ Nijmegen, The Netherlands
[2]Ioffe Institute, 194021 St. Petersburg, Russia

*f.formisano@science.ru.nl

**Abstract**

Excitation, detection, and control of coherent THz magnetic excitation in antiferromagnets are challenging problems that can be addressed using ever shorter laser pulses. We study experimentally excitation of magnetic dynamics at THz frequencies in an antiferromagnetic insulator $CoF_2$ by sub-10 fs laser pulses. Time-resolved pump-probe polarimetric measurements at different temperatures and probe polarizations reveal laser-induced transient circular birefringence oscillating at the frequency of 7.45 THz and present below the Néel temperature. The THz oscillations of circular birefringence are ascribed to oscillations of the magnetic moments of $Co^{2+}$ ions induced by the laser-driven coherent $E_g$ phonon mode via the THz analogue of the transverse piezomagnetic effect. It is also shown that the same pulse launches coherent oscillations of the magnetic linear birefringence at the frequency of 3.4 THz corresponding to the two-magnon mode. Analysis of the probe polarization dependence of the transient magnetic linear birefringence at the frequency of the two-magnon mode enables identifying its symmetry.

**1. Introduction**

The ability to control mechanical, optical, electronic, and magnetic properties of media by light has long intrigued people and fuelled the interest to photo-induced phenomena in condensed matter. This topic became especially appealing after the development of laser sources providing pulses as short as tens and even just a few femtoseconds. Femtosecond all-optical pump-probe experiments opened up a plethora of opportunities for fundamental studies of light-induced dynamics in matter on time-scales of pertinent to elementary motions of atoms [1–4], electrons [3,5–8] or spins [6,9]. It is important to underline that femto- and sub-femtosecond laser pulses represent the shortest stimuli in experimental physics and thus can serve as the fastest ever triggers to speed up electronic [10], photonic [11], and spintronic [12,13] devices.

In this context, antiferromagnets represent the largest and the most intriguing class of magnetically ordered materials. Antiferromagnetic media have found only a few applications but promise to outperform ferromagnetic media in processing and storing information [14,15]. As the frequency of magnetic resonances in antiferromagnets can be up to 100 times higher than in ferromagnetic counterparts and lie in the THz domain, the use of antiferromagnets could drastically increase the rate of processing magnetic bits. Antiferromagnets possess no net magnetic moment and are stable and impervious to external fields. However, for the same reason, control, and detection of magnetic THz excitation in antiferromagnetic media is a challenging problem, which can be tackled by using laser pulses of sub-10 fs duration.

Compensated antiferromagnetic difluorides $MeF_2$, where $Me$=Mn, Fe, Co, have become model materials to investigate novel pathways to manipulate spins in antiferromagnets by short laser pulses [16–18], owing to accessible frequencies of magnetic excitations, strong spin-orbit coupling, and optical transparency. In Ref. [16] impulsive excitation of a two-magnon mode by femtosecond laser pulses has been demonstrated for $MnF_2$ and $FeF_2$. It was suggested that the two-magnon mode is excited via the mechanism of the stimulated Raman scattering (ISRS) [19]. However, mechanisms enabling the detection of two-magnon modes in these and other experiments with antiferromagnets [20,21] have not been discussed in details and remain a matter of debate up to now.

Recently, it has been shown that the excitation of THz optical phonons in $CoF_2$ with mid-infrared laser pulses and subsequent nonlinear lattice dynamics is another way to trigger the magnetic dynamics, manifesting itself in the emergence of transient net magnetization [17]. It has been suggested that a rectified phonon field of a particular symmetry generates a shear strain [22] and thus makes magnetic moments of $Co^{2+}$ ions at different sublattices nonequivalent, in analogy to a static piezomagnetic effect [23–25]. In light of these findings, it is intriguing to verify if a piezomagnetic effect is also feasible at the eigenfrequencies of coherent THz phonons. The THz piezomagnetic effect would extend the range of phenomena where the coherent optical phonons trigger magnetic dynamics [26–29].

Here we report on the experimental study of excitation and detection of THz magnetic dynamics in antiferromagnetic difluoride $CoF_2$ by means of a polarimetric pump-probe technique with sub-10 fs laser pulses. We show that such pulses excite coherent optical phonon mode of the $E_g$ symmetry at the frequency of 7.45 THz. By analyzing how the coherent lattice oscillations perturb the probe pulse polarization, we find that this phonon mode manifests itself through transient linear birefringence, as well as through circular birefringence. We ascribe the latter to the transient induced magnetic moments at $Co^{2+}$ ions due to the local shear distortions associated with the $E_g$ phonon. This can be seen as a high-frequency analogue of the transverse piezomagnetic effect [23]. We further demonstrate excitation of the coherent two-magnon mode at the edge of the Brillouin zone with a frequency of 3.4 THz. Analysis of the probe polarization dependence of the two-magnon contribution to the magnetic linear birefringence enables identification of the symmetry of the two-magnon mode.

The Article is organized as follows. In Section 2, we present a brief overview of the crystal structure, equilibrium lattice, and spin dynamics in $CoF_2$. In Section 3, we discuss the details of the experimental technique. Experimental results and discussion are presented in Section 4. Section 4.1 describes the experimental results on the excitation of the coherent $E_g$ optical phonon mode, which is followed by the discussion of its detection through linear and circular birefringence in Section 4.2, and of the high-frequency transverse piezomagnetic effect in Section 4.3. Experimental results on excitation and detection of the coherent two-magnon mode are presented in Section 4.4. In Section 5, we summarize our findings and discuss further steps towards an understanding of ultrafast magnetic dynamics driven by femtosecond laser pulses.

## 2. Cobalt difluoride $CoF_2$

$CoF_2$ crystallizes in a tetragonal rutile structure [Fig. 1(a)], with the space group $P4_2/mnm$ ($D^{14}_{4h}$, #136, Z=2) [30]. At room temperature, the lattice parameters are $a=b=4.69$ Å and $c=3.19$ Å [30–32]. The unit cell contains two $Co^{2+}$ ions occupying the Wyckoff positions *2a* (0,0,0), and four $F^-$ ions at *4f* positions (0.30346, 0.30346, 0). The fundamental bandgap of $CoF_2$ is ~5.5 eV, as obtained from the transmission spectrum of the studied sample. The absorption in the near-infrared (NIR) range is dominated by the weak *d-d* transitions in $Co^{2+}(3d^7)$ ions [33,34]. Therefore, the excitation of spin and lattice dynamics in this compound by NIR laser pulses is expected to be dominated by non-dissipative mechanisms such as ISRS [19,35].

The unit cell of $CoF_2$ contains six atoms giving rise to 11 optical phonon modes in total [36], with the even $A_{1g}$, $B_{1g}$, $B_{2g}$ and $E_g$ modes being Raman-active. These modes can be driven by NIR laser pulses via the ISRS mechanism [1,19,37]. The $E_g$ mode is of particular importance for piezomagnetism of CoF2 as it is composed of displacements of F$^-$ ions surrounding Co$^{2+}$ ions, as illustrated in Fig. 1(a). Symmetries of other modes are discussed in [39].

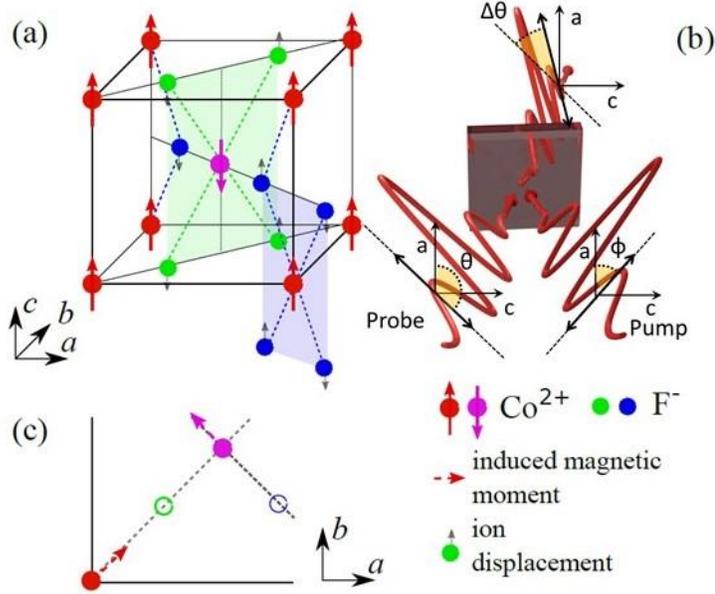

**Figure 1**. (a) Schematic illustration of the crystal and magnetic structure of CoF$_2$, and of ionic displacements associated with the $E_g$ phonon mode in the basis linked to the crystallographic axes following [38]. (b) Experimental geometry. (c) Schematic representation of the phonon-induced magnetic moment via THz transverse piezomagnetic effect (see text). Note that in [59] two $E_g$ modes are presented in the basis rotated by 45 degrees around the c-axis.

The antiferromagnetic phase transition in CoF$_2$ takes place at the Néel temperature $T_N$ = 38 K. Below $T_N$, magnetic moments of Co$^{2+}$ ions form two antiferromagnetically coupled sublattices with magnetizations aligned along the $c$ axis. The effective spin of Co$^{2+}$ ion is $S$ =3/2. Strong spin-orbit coupling in Co$^{2+}$ ions results in a pronounced magnetocrystalline anisotropy. In the electronic structure of the ions, one finds many higher-lying electronic states within the ground state multiplet. The Heisenberg Hamiltonian for Co$^{2+}$ spins in CoF$_2$ without applied magnetic field [40]

$$\mathcal{H} = \sum_{i,j} J_2 S_i \cdot S_j + \frac{1}{2}\sum_{i,i'} J_1 S_i \cdot S_{i'} + \frac{1}{2}\sum_{j,j'} J_3 S_j \cdot S_{j'}$$

$$+g\mu_B \sum_i \{D(S_i^z)^2 - F[(S_i^x)^2 - (S_i^y)^2]\} \qquad (1)$$

includes the dominant intersublattice exchange parameter, $J_2 = 1.6$ meV, and two weaker intrasublattice exchange parameters, $J_1 = -0.15$ meV and $J_3 \sim 0$. Indices $i$ and $j$ denote the sites on different sublattices, $g$ and $\mu_B$ are the $g$-factor and the Bohr magneton, respectively. $D<0$ and $F$ describe uniaxial and in-plane contributions to the magnetic anisotropy, respectively.

Spontaneous Raman scattering experiments showed that in the range 1-6 THz there are several magnetic excitations [41,42]. This range is accessible in pump-probe experiments with NIR femtosecond laser pulses. In particular, antiferromagnetic resonance (AFMR) has a frequency of 1.1 THz [43], two-magnon modes frequency is at 3.4 THz [40], and several magnetic excitons are in the vicinity of 5.8 THz [40]. As discussed in Refs. [40,42], there are three two-magnon modes of $A_{1g}$, $A_{2g}$ and $E_g$ symmetries, having dominating contributions from different points in the Brillouin zone. Their frequencies are primarily governed by the dominating intersublattice exchange $J_2$. Weaker intrasublattice exchange interactions result in an additional splitting of $8\langle S^z \rangle(J_1 - J_3)$ between the $A_{2g}$ and the $E_g$ two-magnon modes. This splitting is related to the difference $4\langle S^z \rangle(J_1 - J_3)$ between the magnon frequencies at the $R$ and the $M$ points of the Brillouin zone and is on the order of 0.07- 0.24 THz [44].

## 3. Experimental

The time-resolved measurements were performed using a degenerative pump-probe setup with a nominally 8 fs-long pump and probe laser pulses, as described in Ref [45]. The pulse spectrum is centered at 1.55 eV and covers the photon energy range from 1.3 to 1.9 eV. The pulses were produced via a compression technique based on spectral broadening of a 35-fs laser pulse in a hollow-core dielectric waveguide (Kaleidoscope™ from Spectra-Physics Vienna) filled with Ne [46]. The 35-fs pulse is produced by a Ti:Sapphire regenerative amplifier at a repetition rate of 1 kHz.

The compressed 8-fs pulse with energy of ~0.5 mJ was split into a pump and probe pulses with an energy per pulse ratio of 3:1 using a custom-made beamsplitter preserving the pulse width. Pump and probe pulses are then cross-polarized using the dual-mirror twisted periscope [47]. The azimuthal angle of the pump polarization ($\phi$) before the incidence on the sample was fixed at $\phi$ =45 deg. The azimuthal angle of the probe polarization $\theta$ was additionally controlled using an achromatic half-waveplate [Fig. 1(b)]. Temporal broadening of the probe pulse caused by the waveplate is of ~15 fs. The angle of incidence for the probe beam was ~ 10 deg, while the pump beam was at normal incidence as shown in Fig. 1(b).

Pump-induced transient changes in the angle of the polarization rotation of the probe $\Delta\theta$ were measured as a function of the pump-probe time delay $t$, controlled by an opto-mechanical translation stage. The polarization rotation angle $\Delta\theta$ at each time delay was detected by means of a custom-made balanced photodetector combined with the Wollaston prism. The pump-probe technique with polarimetric detection enables observation of laser-driven coherent lattice and magnetic excitations [48,49]. More specifically, optically-driven coherent phonons and magnons manifest themselves in modulation of diagonal and off-diagonal components of the dielectric permittivity tensor $\varepsilon_{ij}$ depending on the space-time symmetry of the particular mode. This, in turn, results in modulation of optical birefringence, either linear or circular, and thus affects the polarization of the probe pulses propagating through the sample.

For the experiments, we used a single-crystalline thin plate of $CoF_2$ cut normally to the crystallographic $b$ axis and having a thickness of $d$~500 µm. The single crystal boule of $CoF_2$ was grown by the Bridgman method in platinum crucibles in a helium atmosphere as described in Ref. [50], and the sample was cut from the boule and polished on both sides to the optical quality. The sample was placed in the continuous flow cryostat cooled by liquid helium. Temporal broadening of the pump and probe pulses by the entrance window of the cryostat is estimated to be below 2 fs.

## 4. Results and discussion

### 4.1. Excitation of coherent optical phonon modes in $CoF_2$

Figure 2(a) shows the pump-induced polarization rotation $\Delta\theta$ measured in $CoF_2$ as a function of the pump-probe time delay $t$ at temperatures $T$=20 and 90 K, i.e. below and above $T_N$ =38 K. The incident pump and probe polarizations were $\phi$=45 deg and $\theta$=135 deg with respect to the crystallographic $a$ axis. The durations of pump and probe pulses were of 8 fs because no half-wave plates were used in this set of measurements. As one can see in Fig. 2(a), a strong signal, which is often referred to as a coherent artifact, is present around the pump-probe temporal overlap ($t$=0) followed by a pronounced oscillatory signal, consisting of several superimposed harmonic components. To resolve the frequencies of the oscillatory components in the pump-probe traces, we performed a fast Fourier transform (FFT) analysis of the signals at positive time delays $t > 0.3$ ps.

Figure 2(b) shows the FFT amplitude spectra at $T$=20 and 90 K. Three lines are distinguished in the spectra, with the most pronounced line centered at $f_0$ =7.45 THz and present both below and above $T_N$. Two other lines are centered at 2.0 and 11.1 THz and also present in the whole studied temperature range. The three frequencies are in good agreement with the frequencies of the Raman-active optical phonon modes of $E_g$, $B_{1g}$ and $A_{1g}$ symmetries [see the upper panel in Fig. 2(b)] [40,51]. As expected for phonons, their frequencies do not show pronounced changes when the temperature increased from 10 to 90 K [see inset in Fig. 2(b)]. More particularly, due to spin-phonon coupling the frequency of $E_g$ phonon is expected to decrease by ~0.03 THz as the temperature reaches $T_N$=38 K [52]. Such a frequency shift is simply below the spectral resolution of the measurements, which we estimate to be of ~0.1 THz. In our work, the frequencies are deduced from the FFT of the measured dynamics at a sub-10 ps timescale. A better frequency resolution of the measurements would require measurements in a broader time window [51,52].

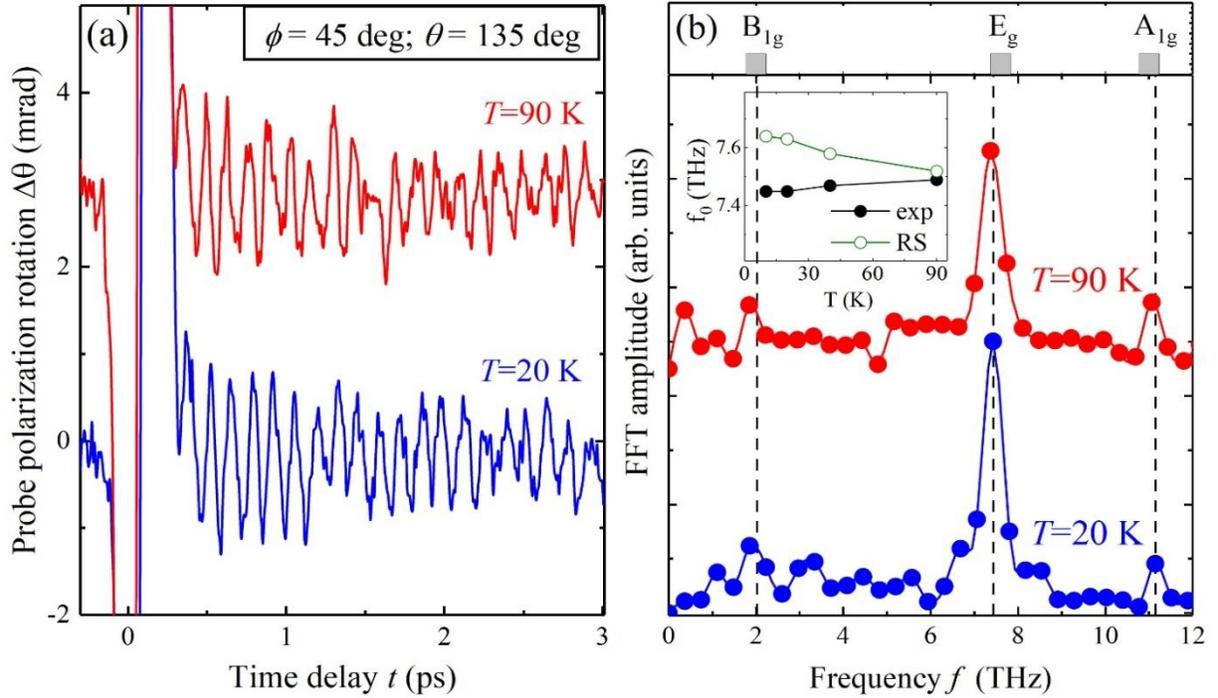

**Figure 2.** (a) Probe polarization rotation $\Delta\theta$ as a function of the pump-probe delay $t$ at $T$=20 K (blue curve) and $T$=90 K (red curve) for $\phi$=45 deg and $\theta$=135 deg. (b) FFT spectra of the pump-probe traces in (a) at $T$=20 K (blue curve) and $T$=90 K (red curve). Upper panel in (b) shows the frequencies of optical phonon modes as obtained from spontaneous Raman scattering (RS) [40, 51]. Inset in (b) shows the temperature dependence of the central frequency $f_0$ of the $E_g$ mode as obtained in the pump-probe experiments (closed symbols) and from RS [40, 51] (open symbols). Lines are the guides for an eye.

Therefore, in agreement with the previously published data, we show that sub-10 fs laser pulses can excite $CoF_2$ coherent optical phonon modes. ISRS is the most plausible mechanism of the phonon modes excitation [19,53], as discussed in Suppl. Mater. [39]. However, the mechanisms allowing detection of the $E_g$ phonon mode in our experiments require detailed analysis, as we show below.

*4.2. Detection of the $E_g$ phonon mode in $CoF_2$*

In order to reveal the mechanisms responsible for the detection of the coherent $E_g$ phonon mode, we have studied the dependence of the oscillations of $\Delta\theta(t)$ at $f_0$=7.45 THz on the incoming probe polarization $\theta$. Figure 3(a) shows the pump-probe traces obtained at $T$=20 K with incoming probe pulses linearly polarized at $\theta$=45, 90, and 135 deg. The incoming pump polarization was $\phi$=45 deg, which ensures efficient excitation of the $E_g$ mode via ISRS. As one can see, the amplitude of the oscillations at $f_0$=7.45 THz is dependent on the incoming polarization of the probe pulse, $\theta$. First, the amplitude of the probe polarization oscillations decreases noticeably when the probe pulse is initially polarized at $\theta$=90 deg, i.e. along the $c$ axis. Second, when the incoming probe polarization changes from $\theta$=45 to 135 deg, the initial phase of the oscillations of $\Delta\theta(t)$ at $f_0$=7.45 THz changes by π, i.e. the amplitude $\Delta\theta_0$ changes its sign.

Figures 3(c-d) present the dependences of the signed amplitude $\Delta\theta_0$ of the 7.45 THz component on the probe polarization angle $\theta$ at $T$=10, 20 and 90 K. Since the pump-probe traces [Fig. 3(a)] also contain contributions at frequencies 2.0 and 11.1 THz, the amplitude $\Delta\theta_0$ was obtained by fitting the line at $f_0$=7.45 THz in the FFT spectrum of the pump-probe traces using the Lorentz line shape [see Fig. 3(b)]. The change of the initial phase of the oscillations by $\pi$ is taken into account as the sign change of $\Delta\theta_0(\theta)$. As one can see, in Figs. 3(c-e) the dependence $\Delta\theta_0(\theta)$ below and above $T_N$ can be approximated with a sum of probe-polarization dependent and probe-polarization independent contributions with the amplitudes $\Delta\theta_A$ and $\Delta\theta_B$, respectively.

$$\Delta\theta_0(\theta) = \Delta\theta_A \sin(2\theta + \delta) + \Delta\theta_B, \qquad (2)$$

where $\delta=\pi/36$ is the phase shift. Figures 3(c-e) show that the probe-polarization dependent contribution is minimal when the probe, $\theta$, is initially polarized close to the *a*- or *c*-axis. The small phase shift $\delta$ of 5 deg in the fit function is due to a slight misalignment of the crystal axes with respect to the laboratory frame. Nominally probe-polarization independent contribution $\Delta\theta_B$ is introduced in Eq. (2) as the simplest way to account for the systematic negative shift of $\Delta\theta_0$, seen at both 45<$\theta$<90 deg and 90<$\theta$<180 deg. Goodness of fit shown in Figs. 3 (c-e) was estimated accepting the fit function $\Delta\theta_0(\theta) = \Delta\theta_A \sin(2\theta + \delta) + \Delta\theta_B$ as the null hypothesis and calculating adjusted $R^2$-values ($R_{adj}^2$) and reduced $\chi^2$-values as explained in Ref. [54]. We found that $R_{adj}^2$ values exceed 0.75 and reduced $\chi^2$-values are below $10^{-7}$. According to Table III from Ref. [54], these numbers show that the null hypothesis can be accepted.

The data in Figs. 3(f, g) show that the probe-polarization dependent contribution $\Delta\theta_A$ does not experience pronounced systematic changes as the temperature increases above $T_N$. In contrast, the polarization-independent contribution $\Delta\theta_B$ decreases upon temperature increase and vanishes above the Néel point, $T > T_N$. Different polarization and temperature dependences of the two contributions to $\Delta\theta_0(\theta)$ suggest that the detection of the laser-driven coherent $E_g$ mode is realized via two mechanisms. The probe-polarization dependent transient rotation $\Delta\theta(t) = [\Delta\theta_A \sin(2\theta + \delta)]\sin(2\pi f_0 t)$ is typical for the transient modulation of the *linear* birefringence by a coherent phonon, as was demonstrated previously in several materials [48,53,55]. The probe polarization-independent rotation $\Delta\theta(t) = \Delta\theta_B \sin(2\pi f_0 t)$, induced by a coherent phonon, is less trivial and must be assigned to the *circular* birefringence. We note that in an anisotropic medium, lattice vibrations changing linear birefringence can also result in a polarization rotation $\Delta\theta_0$ independent on the initial orientation of the probe polarization $\theta$. However, there is no reason to expect that this effect is different below and above the Néel point as shown in Fig. 3(g) (see also Fig. S1 in Suppl. Mater. [39]).

To describe the observed the probe-polarization and temperature dependencies of the polarization rotation $\Delta\theta(t)$, we recall that the linear and circular birefringence originate from symmetric $\varepsilon_{ij}^{(s)} = \varepsilon_{ji}^{(s)}$ and antisymmetric $\varepsilon_{ij}^{(a)} = -\varepsilon_{ji}^{(a)}$ parts of the dielectric tensor, respectively. If we assume a linear dependence between the dielectric tensor $\varepsilon_{ij}(t)$ and the phonon normal coordinate $Q(t)$, an excitation of a coherent phonon mode results in a contribution to $\varepsilon_{ij}^{(s)}$ which depends on the pump-probe delay $t$

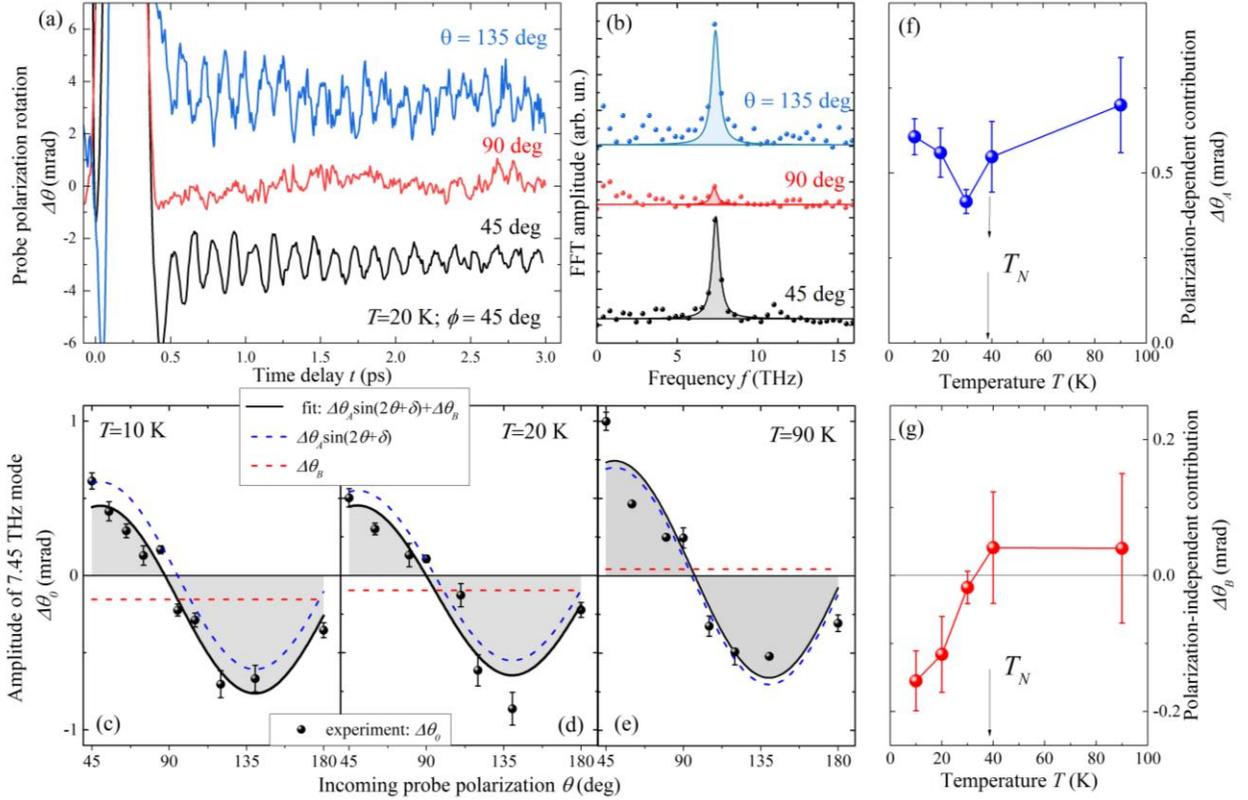

**Figure 3**. (a) Probe polarization rotation $\Delta\theta$ as a function of the pump-probe delay $t$ at $T=20$ K for the incoming pump $\phi=45$ deg and probe polarizations $\theta=45, 90, 135$ deg. (b) FFT spectra of the pump-probe traces shown in the panel (a). (c-e) Amplitude $\Delta\theta_0$ of the feature at 7.45 THz (symbols) as a function of probe polarization, $\theta$, at $T=10, 20,$ and 90 K. $\phi=45$ deg in all panels. Black solid line is the fit of the data to the function $\Delta\theta_0 = \Delta\theta_A \sin(2\theta + \delta) + \Delta\theta_B$, using the damped least-squares method, with $\delta=\pi/36$ being the phase shift. Blue and red dashed lines show the probe polarization dependent $\Delta\theta_A \sin(2\theta + \delta)$ and probe polarization independent $\Delta\theta_B$ contributions, respectively. (f,g) Fit parameters $\Delta\theta_A$ and $\Delta\theta_B$ as a function of the sample temperature. Error bars in Figs. 3 (c-e) show the standard deviation of the set of the points in the FFT spectra in the range of 20-70 THz where no peaks associated with phonon modes are present. Error bars in Figs. 3 (g, f) represent the goodness of the fit of $\Delta\theta_0$ using Eq. (2).

$$\delta\varepsilon_{ij}^{(s)}(t) = R_{ij}Q(t), \qquad (3)$$

where $R_{ij}$ is the Raman tensor for the particular phonon mode. In the case of the $E_g$ phonon, the nonzero Raman tensor components are $R_{ac} = R_{ca}$ [34,39,55]. Hence, for the symmetric part of the dielectric permittivity tensor, we obtain

$$\varepsilon_{ij}^{(s)}(t) = \begin{pmatrix} \varepsilon_{aa} & 0 & R_{ac}Q(t) \\ 0 & \varepsilon_{aa} & 0 \\ R_{ca}Q(t) & 0 & \varepsilon_{cc} \end{pmatrix}, \qquad (4)$$

where $\varepsilon_{aa}^{(s)}$, and $\varepsilon_{cc}^{(s)}$ are the components of the dielectric permittivity of the unperturbed medium. Both $\varepsilon_{aa(cc)}^{(s)}$ and $R_{ac}$ are real in the non-dissipative regime of light-matter interactions. As a result, $\varepsilon_{ij}^{(s)}(t)$ gives rise to the transient ellipticity of the probe pulses propagating along the $b$ axis, which occurs at the frequency $f_0$ of the $E_g$ phonon mode [48,49]. Static birefringence of the crystal $\sqrt{\varepsilon_{aa}^{(s)}} - \sqrt{\varepsilon_{cc}^{(s)}}$ results in conversion of the transient ellipticity to rotation upon probe propagation which is detected in our experiments, as discussed in Suppl. Mater [39], Sec. 2.1:

$$\Delta\theta(t,d) = \frac{2R_{ac}Q(t)}{\varepsilon_{aa}^{(s)} - \varepsilon_{cc}^{(s)}} \left[ 1 - \cos\frac{2\pi\Delta nd}{\lambda} \left[ \cos^2 2\theta + \left( \cos\frac{2\pi\Delta nd}{\lambda} \sin 2\theta \right)^2 \right]^{-1} \right]. \qquad (5)$$

Conversion of probe ellipticity to rotation is more pronounced when the incoming probe polarization is at $\theta=45$ deg to the crystallographic axes, as compared to the case when $\theta=0$ deg. This is in agreement with the experimentally observed dependence approximated for a sake of simplicity by the function $\Delta\theta_A \sin(2\theta + \delta)$ [Fig. 3(c-e)]. Such a contribution is not expected to show a pronounced temperature dependence and must be present above and below $T_N$. Weak changes of $\Delta\theta_A$ with temperature, seen in Fig. 3(g), may be ascribed to temperature-dependent crystallographic linear birefringence $\sqrt{\varepsilon_{aa}^{(s)}} - \sqrt{\varepsilon_{cc}^{(s)}}$ [56], as well as to temperature variations of Raman tensor [42].

For the antisymmetric part of the dielectric permittivity tensor, which gives rise to circular birefringence detected in our experiment for the probe pulse propagating along the $b$ axis, we propose that

$$\varepsilon_{ij}^{(a)}(t) = \begin{pmatrix} 0 & 0 & iPQ(t) \\ 0 & 0 & 0 \\ -iPQ(t) & 0 & 0 \end{pmatrix}, \quad (6)$$

where $P$ is a phenomenological parameter. In this case, the probe polarization will acquire transient rotation upon propagation through the sample along the $b$ axis (see Suppl. Mater [39], Sec. 2.2). The transient rotation will be proportional to the phonon coordinate. If one neglects linear birefringence in the isotropic crystal, the rotation is independent of the probe incoming polarization, $\theta$:

$$\Delta\theta_B \sin(2\pi f_0 t) \sim \frac{\sqrt{2}\pi d}{\lambda\sqrt{\varepsilon_{aa}^{(s)} + \varepsilon_{cc}^{(s)}}} PQ(t). \quad (7)$$

If the birefringence cannot be neglected, the transient rotation will be larger when $\theta$=45, 135 deg as compared to $\theta$=90, 180 deg (see Eq. (S12) and related text in Suppl. Mater [39] Sec. 2.2). This can indeed be seen in the experimental data if one plots the dependence of $\Delta\theta_0$ on temperature at different incoming probe polarizations, $\theta$, (see Suppl. Mater [39] Sec. 2.2 and Fig. S1 therein).

Importantly, the probe-polarization independent contribution $\Delta\theta_B$ to the detected signal is present only below the $T_N$ [Fig. 3(g)]. Since the coherent phonon mode also persists above $T_N$, this implies that the coefficient $P$ is temperature-dependent. Therefore, one can argue that the emergence of $\varepsilon_{ij}^{(a)}(t)$ [Eq. (6)] suggests that the lattice vibrations break the symmetry of the medium in the very same way as a magnetization $M_b$ along the $b$ axis [58]. Assuming that the probe pulse excites only electronic transitions, the magnetization may have either spin or orbital origin.

*4.3. THz piezomagnetic effect in CoF₂*

In order to elucidate the origin of the transient circular birefringence and the feasibility of the phonon-induced magnetic moment oscillating at the THz frequency, we recall that below $T_N$, the crystallographic and magnetic symmetry of $CoF_2$ allows a transverse piezomagnetic effect [23,25]. It means that under shear strains $\Lambda_{(ac)bc}$ applied in the $ac$ or $bc$ plane containing equilibrium antiferromagnetic vector, the crystal can acquire a net magnetization along the $b$ or $a$ axis, respectively [23]. Specifically, under static strain, this effect manifests itself in a deviation of the antiferromagnetic vector from the $c$ axis in the $ac(bc)$ plane and a canting of the magnetic moments of the antiferromagnetic sublattices, which results in net magnetization along the $b(a)$ axis [see Fig. 1(c)].

Excitation of the coherent phonon mode of the $E_g$ symmetry gives rise to the local lattice distortions in the $ac$ and $bc$ planes, as shown in Fig. 1(c). These shear distortions for $Co^{2+}$ ions at two sublattices emerge in two orthogonal planes [Fig. 1 (a)] and are $\Lambda_{ac}(t) = \Lambda_{bc}(t) \sim Q(t)$ and $\Lambda_{ac}(t) = -\Lambda_{bc}(t) \sim Q(t)$, respectively [59]. Note, that this dynamic strain is different from a static shear strain in the $ac$ or $bc$ planes. The latter would have the same sign for $Co^{2+}$ ions at both sublattices. Thus, to derive the effect of the $E_g$ phonon on the magnetic structure, we write down a modified expression for the free energy in the presence of the shear strains $\Lambda_{ac}, \Lambda_{bc}$ and $\Lambda_{ac}, -\Lambda_{bc}$ for the $Co^{2+}$ ions at two sublattices (see Suppl. Mater. [39], Section 3 for details):

$$F = \tfrac{1}{2}A_1(l_a^2 + l_b^2) + \tfrac{1}{2}A_2 \boldsymbol{m}^2 + \tfrac{1}{2}d_1 m_c^2 + d_2(m_a l_b + m_b l_a)$$
$$+\lambda_1(l_a \Lambda_{bc}(t) + m_b \Lambda_{ac}(t))l_c + \eta_1(l_a \Lambda_{ac}(t) + m_b \Lambda_{bc}(t))l_c, \qquad (8)$$

where $\mathbf{L} = \mathbf{M}_1 - \mathbf{M}_2$, $\mathbf{l}_i = \mathbf{L}_i/L$, $\mathbf{M} = \mathbf{M}_1 + \mathbf{M}_2$, $\mathbf{m}_i = \mathbf{M}_i/M$, with subscript $i=a,b,c$ specifying the projections of vectors $\mathbf{L}, \mathbf{M}, \mathbf{l}, \mathbf{m}$ on the axes. Parameters $A_1, A_2$ account for the exchange interaction, $d_1$ and $d_2$ for the magnetocrystalline anisotropy, and the Dzyaloshinskii-Moriya interaction, while $\lambda_1$ and $\eta_1$ are the piezomagnetic coefficients [23]. Minimizing the free energy for nonzero $\Lambda_{(ac)bc}$ one gets that the local shear distortions, associated with the coherent $E_g$ mode, induce magnetic moments $\delta\mathbf{m}_{1,2}(t)$ at the two $Co^{2+}$ sublattices, and the net magnetization $\delta\mathbf{m}(t)$:

$$\delta\mathbf{m}(t) = \left(\frac{d_2\lambda_1 - A_1\eta_1}{A_1 A_2 - d_2^2}\Lambda_{bc} - \frac{d_2\eta_1 - A_1\lambda_1}{A_1 A_2 - d_2^2}\Lambda_{ac}\right)\mathbf{b}, \qquad (9)$$

where $\mathbf{b}$ is the unit vectors along the $b$ axis [Fig. 1(c)]. Hence coherent phonon corresponding to the $E_g$ mode generates a net magnetic moment in the $ab$ plane $\delta\mathbf{m}(t)$ oscillating at the frequency $f_0$ of the phonon. The induced magnetic moment has, in particular, a non-zero component along the $b$ axis, and thus can be seen in circular birefringence of light propagating along the $b$ axis $\varepsilon_{ac}^{(a)}(t) \sim i\delta m_b(t)$, in agreement with the experimental observations. We note that the piezomagnetic effect is the odd effect with respect to reversal of the antiferromagnetic vector [23]. Thus, presence of 180°-antiferromagnetic domains may lead to partial quenching of the induced oscillating magnetic moment and account for weakness of the detected signal $\Delta\theta_B$.

The magnetic moment of $Co^{2+}$ ion in $CoF_2$ is a sum of spin and orbital moments. In $CoF_2$, the frequency $f_0$=7.45 THz of the $E_g$ mode is considerably higher than the highest frequency 1.1 THz of the antiferromagnetic resonance [43], and the observed high-frequency magnetic dynamics cannot be due to conventional dynamics of spins. To understand the origin of the magnetization associated with the $E_g$ phonon one should consider the dynamics of orbital states as well. The magnetism of the $E_g$ phonon mode and its effect on spin excitations in $CoF_2$ was analyzed in Ref. [59] taking into account mixing of excited orbital states into the ground doublet, shear distortion of the crystal field with the $E_g$ symmetry, and the first-order mixing of the magnetic excitations with the phonon mode. It was shown that, in the case of noncoherent $E_g$ phonons, resulting phonon-induced high-frequency transverse spin polarization should manifest itself through infrared magnetic-dipole absorption at the frequency of this phonon, indeed reported in the same work [59]. In our experiments, we excite *coherent* $E_g$ phonon with uniform initial phase of shear ionic displacements within the excited volume of material. This also makes the phonon-induced spin polarization and its oscillations coherent at all involved $Co^{2+}$ sites. As a result, the induced net magnetic moment is detected through the macroscopic magnetic circular birefringence.

We note that the symmetry consideration presented above shows that the coherent phonon mode of the $E_g$ symmetry can, indeed, contribute to the THz analogue of the piezomagnetic effect by introducing local shear distortions. However, the direction and proportionality coefficient of the induced THz magnetic moment differs from the one observed when applying conventional static shear strain [23] or transverse acoustic waves [60]. This is a result of the difference in local distortions induced by static strain or acoustic phonons as compared to the distortions associated with optical phonons.

It is also worth noting that, recently, the longitudinal piezomagnetic effect, associated with the rectified phonon field of the $B_{2g}$ symmetry, has been reported in $CoF_2$ [17]. In that case, the quasi-static shear distortions were realized due to nonlinearities of phononic excitations and the orbital moment along the *c* axis was induced. The local distortions of the $B_{2g}$ symmetry resemble the ones obtained with the static shear strain in the *ab* plane, only if the displacements of two $F^-$ ions, closest to each of $Co^{2+}$ ions, in positions *2a* and *4f* are considered. In principle, one cannot exclude that the magnetization induced by the dynamic shear strain from phonons is different from the one induced by a static share strain if the effect of all 6F⁻ ions displacement in the elementary cell is taken into account.

*4.4. Excitation and detection of the two-magnon mode in CoF$_2$*

We now turn to the discussion of the excitation and detection of the two-magnon mode by sub-10 fs laser pulses. In CoF$_2$, the two-magnon mode has a frequency of 3.4 THz [40]. It is known that, in the spontaneous Raman spectra, the two-magnon mode is considerably weaker than the $E_g$ phonon mode [40]. Therefore, in time-resolved pump-probe experiments, the contribution associated with the two-magnon mode is masked by the much stronger contribution from the $E_g$ phonon. Indeed, the pump-probe time traces [Figs. 2(a) and 3(a)] and their FFTs are dominated by the 7.45 THz component. Hence, we first examined if the two-magnon mode can be resolved in the pump-probe signal.

Figure 4 shows the FFT spectra of the pump-probe traces obtained at $\theta$=45, 90 deg at two temperatures $T$=10, 20 K below, and at $T$=60 K above the Néel temperature. At temperatures below $T_N$ [Fig. 4(a,b)], a peak with a width of 0.3±0.1THz is seen in the vicinity of $f_{2M}$ =3.4 THz. The latter frequency corresponds to the one of the two-magnon modes and can be resolved when $\theta$=90 deg. We note that the peak centered at 3.4 THz seen also when $\theta$=45 deg and $\phi$=45 deg does not exceed other features in the range of 1.5-6.5 THz, which defines the noise level. This feature disappears above $T_N$ for all studied polarizations [Fig. 4(c)], suggesting that it can be assigned to the laser-induced coherent two-magnon mode.

As follows from Fig. 4(a,b), probing the laser-induced two-magnon mode is sensitive to the incoming probe polarization $\theta$. Rotation of the pump polarization by 45 deg, in contrast, has no significant effect on the excitation efficiency, as can be seen from comparing the spectra obtained for the same probe polarization $\theta$=90 deg and two different pump polarizations $\phi$=45 and 90 deg (see Fig. 4(a)). This finding suggests that the coherent two-magnon mode is excited independently from the coherent phonon mode at 7.45 THz.

To analyze the detection of the two-magnon mode by femtosecond laser pulses, we follow the same approach as for the analysis of the $E_g$ phonon in Sec. 4.2. We analyze which component of the dielectric tensor is modulated by a particular coherent mode, based on the properties of the corresponding Raman tensor [Eq. (2)]. As discussed in Sec. 2, the two-magnon peak in Raman spectra of $CoF_2$ contains three modes of $A_{1g}$, $A_{2g}$, and $E_g$ symmetries [42]. The corresponding Raman tensors components are $R^{2M}_{aa}, R^{2M}_{bb}, R^{2M}_{cc}$, $R^{2M}_{ab} = R^{2M}_{ba}$, and $R^{2M}_{ac} = R^{2M}_{ca}$, $R^{2M}_{bc} = R^{2M}_{cb}$, respectively. In the experimental geometry with probe pulses propagating close to the *b* axis, only the two-magnon modes of $A_{1g}$ and $E_g$ symmetries are expected to be detected. The corresponding modulation of the dielectric tensor is $\delta\varepsilon^{(s)}_{aa}(t), \delta\varepsilon^{(s)}_{cc}(t)$ for the $A_{1g}$ mode, and $\delta\varepsilon^{(s)}_{ac}(t) = \delta\varepsilon^{(s)}_{ca}(t)$ for the $E_g$ mode. Assuming that the Raman tensor components can be complex, in general, we derive that the two-magnon mode of the $E_g$ symmetry induces transient probe polarization rotation

$$\Delta\theta \sin(2\pi f_{2M} t) \sim \frac{\pi d}{\lambda} \frac{2R_{ac}^{2M} \cos(2\theta)}{\sqrt{\varepsilon_{cc}^{(s)}} - \sqrt{\varepsilon_{aa}^{(s)}}} Q(t), \tag{10}$$

which agrees with the experimental data in Fig.4(a,b) showing that the laser-induced two-magnon mode is reliably detected when $\theta=90$ deg. Thus, the probe polarization dependence of the signal suggests that the detected two-magnon mode is of $E_g$ symmetry and implies a dominating contribution of magnons at the $R$ point in the Brillouin zone [40,42].

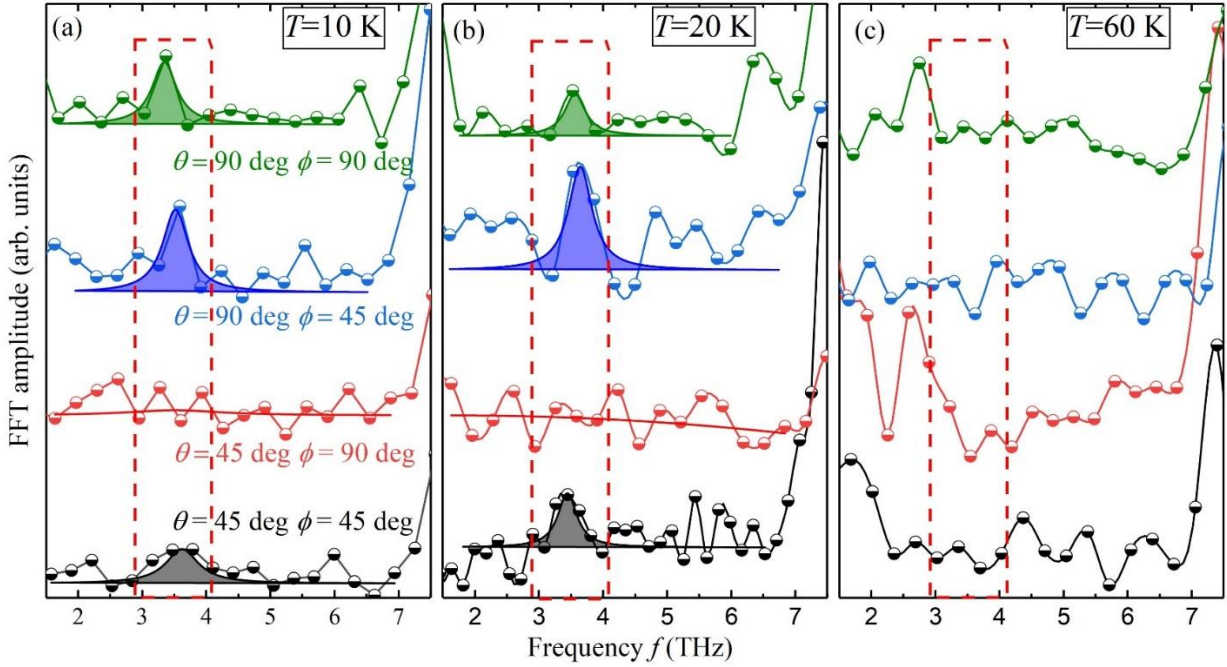

**Fig. 4**. Excitation and detection of the coherent two-magnon mode in $CoF_2$. FFT spectra of the pump-probe traces $\Delta\theta(t)$ measured at four combinations of the pump, $\phi$=45,90 deg, and probe, $\theta$=45,90 deg, polarizations with respect to the *a* axis, at (a) *T*=10, (b) 20, and (b) 60 K. Dashed frames show the range corresponding to the two-magnon mode around $f_{2M}$=3.4 THz. Thick lines show the Lorenz fit of the feature at $f_{2M}$=3.4 THz with the corresponded color filled area.

Theory for the Raman scattering from the two-magnon mode was developed in Refs. [40,42]. The Hamiltonian for the two-magnon light scattering in $CoF_2$ has a form:

$$\mathcal{H}_{int} = \frac{1}{2}\sum_{r,\delta}\varphi(\delta)(S_r^+ S_{r+\delta}^- + S_r^- S_{r+\delta}^+ + \gamma S_r^z S_{r+\delta}^z)$$

$$+ \frac{1}{2}\sum_{r,\delta}\varphi'(\delta)(S_r^+ S_{r+\delta}^- - S_r^- S_{r+\delta}^+), \tag{11}$$

where $r$ runs over all the spin sites, $\delta$ connects the sites with their next-nearest neighbors on the opposite sublattice, and $\gamma$ is a weighting factor. The last term in Eq. (11) emerges due to the large magnetic anisotropy field in $CoF_2$ [50,61,62]. Taking into account that the probe pulse propagates close to the $b$ axis, the expressions for the symmetry terms $\varphi(\delta)$ and $\varphi'(\delta)$ in Eq. (11) are reduced to:

$$\varphi(\delta) = B_1 e_1^a e_2^a + B_2 e_1^c e_2^c + B_4(e_1^a e_2^c + e_1^c e_2^a)\sigma_\delta^a \sigma_\delta^c + B_5(e_1^a e_2^c - e_1^c e_2^a)\sigma_\delta^a \sigma_\delta^c;$$

$$\varphi'(\delta) = B_6 e_1^a e_2^a, \tag{12}$$

where $B_{1,...,6}$ are the coefficients of magneto-optical coupling, $\mathbf{e}_1$ and $\mathbf{e}_2$ are normalized components of the electric field of light, $e_{1,2}^i$ are their projection on the $i$ axis, and $\boldsymbol{\sigma}_\delta = \text{sign}(\boldsymbol{\delta})$. The contribution to the permittivity tensor $\varepsilon_{ij}^{(s)}(\mathbf{r})$ originating from two spin operators at different sublattices is:

$$\varepsilon_{ij}^{(s)}(\mathbf{r}) = \sum_\delta \sum_{kl} B_{ijkl}(\mathbf{r}) S_r^k S_{r+\delta}^l, \tag{13}$$

and describes the modulation of the dielectric permittivity tensor by coherent two-magnon mode. $B_{ijkl}(\mathbf{r})$ corresponds to various magneto-optical coefficients $B_{1,...,6}$ entering in Eq. (13).

The strength of modulation of the dielectric permittivity tensor components is defined by the particular coefficient $B_{1,...,6}$ in Eqs. (12,13). $B_4$ and $B_5$ account for modulation of the off-diagonal components of the dielectric permittivity tensor and must be assigned to the two-magnon mode of the $E_g$ symmetry, while $B_1$, $B_2$, and $B_6$ account for modulation of the diagonal components by the mode of $A_{1g}$ symmetry. In Ref. [42], based on a detailed analysis of the two-magnon lines in the spontaneous Raman spectra of $CoF_2$, the ratios between values of $B_{1,...,6}$ were estimated. In particular, it was obtained that $|B_1|/|B_4 \pm B_5| \sim 0.3$, $|B_2|/|B_4 \pm B_5| \sim 0.8$, which suggests that the contribution from the $A_{1g}$ mode to the modulation of $\varepsilon_{ij}$ should be weaker than that from the $E_g$ mode. This agrees with the experimental observation, where the laser-driven two-magnon mode manifests itself in the pump-probe signal obtained with the probe polarized at $\theta=90$ deg [Fig. 4(a,b)].

## 5. Conclusions and outlook

Sub-10 fs near-infrared laser pulses efficiently launch in antiferromagnetic $CoF_2$ two THz coherent excitations of magnetic origin. It is shown that laser-induced oscillations of ions in the crystal lattice corresponding to the $E_g$ phonon result not only in a linear but also circular birefringence for the probe pulse propagating through the crystal. The circular birefringence is explained as a result of net magnetic moment induced by local distortions, associated with the lattice vibration, in combination with the transverse piezomagnetic effect. The very high frequency of the THz phonons implies that the net dynamic magnetic moment must originate from the crystal field distortions having an impact on the mixing of excited and ground orbital states of $Co^{2+}$ ions. For noncoherent phonons, such effect has been shown [59] to manifest itself through the infrared magnetic-dipole absorption at the $E_g$ phonon frequency, while the laser-driven coherent phonon results in macroscopic magnetic dynamics measurable through magneto-optical effect in the visible range. Thus, transverse piezomagnetism at frequencies of THz phonons opens up a novel possibility to access magnetic dynamics at such frequencies and extends a range of recently explored coherent phononic effects in ultrafast magnetism [17,28,29,63].

We note that in $CoF_2$ the transverse piezomagnetic effect at the frequency of the $E_g$ phonon can be much smaller than the static one which has a substantial contribution of spin rotations in the antiferromagnetic sublattices. Therefore, it is interesting to explore the high-frequency transverse piezomagnetic effect in a material with a phonon mode frequency close to the frequency of the magnetic resonance, where the piezomagnetic effect could trigger coupled spin-orbital eigenmodes [64].

Eventually, we have detected the laser-induced 3.4 THz coherent two-magnon mode. By analyzing the sensitivity of the transient probe polarization rotation, $\Delta\theta$, at this frequency to the incoming probe polarization, $\theta$, we determine that the two-magnon mode detected in the experiment is of $E_g$ symmetry. It points out that the dominating contribution comes from magnons at the *R* point in the Brillouin zone. Thus, the polarimetric detection of the coherent two-magnon mode appears to provide complementary information with respect to other methods employed earlier [16,18].

**Acknowledgments**

The authors thank S. Semin, C. M. Berkhout for technical support and M. A. Prosnikov for fruitful discussions. The work was supported by the Nederlandse Organisatie voor Wetenschappelijk Onderzoek (NWO). A.M.K. acknowledges support from Russian Science Foundation, grant No. 20-42-04405. R.M.D. and R.V.P acknowledge joint support from the Russian Foundation for Basic Research, grant No. 19-52-12065 and Deutsche Forschungsgemeinschaft, project TRR 160 ICRC, B9. The collaboration between Radboud University and the Ioffe Institute is a part of the EU COST Action CA17123 Magnetofon.

**References**

[1] Yan Y. X., Gamble E. B., and Nelson K. A. 1985 *J. Chem. Phys.* **83** 5391.

[2] Matsuda O., M. Larciprete C., Li Voti R., and Wright O. B. *Ultrasonics* 2015 **56** 3.

[3] Wegkamp D. and Stähler J. 2015 *Prog. Surf. Sci.* **90** 464.

[4] Mankowsky R., Von Hoegen A., Först M., and Cavalleri A. 2017 *Phys. Rev. Lett.* **118**, 197601.

[5] Garg M., Zhan M., Luu T. T., Lakhotia H., Klostermann T., A. Guggenmos, and Goulielmakis E. 2016 *Nature* **538** 359.

[6] Siegrist F., Gessner J. A., Ossiander M., Denker C., Chang Y.-P., Schröder M. C., Guggenmos A., Cui Y., Walowski J., Martens U., Dewhurst J. K., Kleineberg U., Münzenberg M., Sharma S., and Schultze M. 2019 *Nature* **571** 240.

[7] Fausti D., Tobey R. I., Dean N., Kaiser S., Dienst A., Hoffmann M. C., Pyon S., Takayama T., Takagi H., and Cavalleri A. 2011 *Science* **331** 189.

[8] de Jong S., Kukreja R., Trabant C., Pontius N., Chang C. F., Kachel T., Beye M., Sorgenfrei F., Back C. H., Bräuer B., *et al.* 2013 *Nat. Mater.* **12** 882.

[9] Kirilyuk A., Kimel A. V., and Rasing Th. 2010 *Rev. Mod. Phys.* **82** 2731.

[10] Karnetzky C., Zimmermann P., Trummer C., Duque Sierra C., Wörle M., Kienberger R., and Holleitner A. 2018 *Nat. Commun.* **9** 2471.

[11] Ono M., Hata M., Tsunekawa M., Nozaki K., Sumikura H., Chiba H., and Notomi M. 2020 *Nat. Photon.* **14** 37.

[12] Walowski J. and Münzenberg M. 2016 *J. Appl. Phys.* **120** 140901.

[13] Kimel A. V. and Li M. 2019 *Nat. Rev. Mater.* **4** 189.

[14] Jungwirth T., Marti X., Wadley P., and Wunderlich J. 2016 *Nat. Nanotechnol.* **11** 231.

[15] Baltz V., Manchon A., Tsoi M., Moriyama T., Ono T., and Tserkovnyak Y. 2018 *Rev. Mod. Phys.* **90** 015005.

[16] Zhao J., Bragas A. V., Merlin R., and Lockwood D. J., 2006 *Phys. Rev. B* **73** 184434.

[17] Disa A. S., Fechner M., Nova T. F., Liu B., Först M., Prabhakaran D., Radaelli P. G., and Cavalleri A. 2020 *Nat. Phys.* **16**, 937 (2020).

[18] Zhao J., Bragas A. V., Lockwood D. J., and Merlin R. 2004 *Phys. Rev. Lett.* **93** 107203.

[19] Merlin R. 1997 *Solid State Commun.* **102** 207.

[20] Bossini D., Dal Conte S., Cerullo G., Gomonay O., Pisarev R. V., Borovsak M., Mihailovic D., Sinova J., Mentink J. H., Rasing Th., and Kimel A. V., 2019 *Phys. Rev. B* **100** 024428.

[21] Bossini D., Dal Conte S., Hashimoto Y., Secchi A., Pisarev R. V., Rasing Th., Cerullo G., and Kimel A. V., 2016 *Nat. Commun.* **7** 106045.

[22] Radaelli P. G. 2018 *Phys. Rev. B* **97** 085145.

[23] Borovik-Romanov A. S. 1960 *J. Exptl. Theor. Phys.* **38** 1088 [*Sov. Phys. - JETP* **11** 786].

[24] Dzialoshinskii I. E. 1958 *J. Exptl. Theor. Phys.* **33** 807 [*Sov. Phys. - JETP* **6** 621].

[25] Moriya T. 1959 *J. Phys. Chem. Solids* **11** 73.

[26] Rebane Y. T. 1983 *Zh. Eksp. Teor. Fiz.* **84** 2323 [*Sov. Phys. - JETP* **57** 1356].

[27] Juraschek D. M., Fechner M., Balatsky A. V., and Spaldin N. A. 2017 *Phys. Rev. Mater.* **1** 14401.

[28] Stupakiewicz A., Davies C. S., Szerenos K., Afanasiev D., Rabinovich K. S., Boris A. V., Caviglia A., Kimel A. V., and Kirilyuk A. 2021 *Nat. Phys.* **17** 489.

[29] Nova T. F., Cartella A., Cantaluppi A., Först M., Bossini D., Mikhaylovskiy R. V., Kimel A. V., Merlin R., and Cavalleri A. 2017 *Nat. Phys.* **13** 132.

[30] Stout J. W. and Reed S. A. 1954 *J. Am. Chem. Soc.* **76** 5279.

[31] Costa M. M. R., Paixão J. A., De Almeida M. J. M., and Andrade L. C. R. 1993 *Acta Crystallogr. B* **49** 591.

[32] Belyaeva A. I., Eremenko V. V., Mikhailov N. N., and Petrov S. V. 1966 *J. Exptl. Theor. Phys.* **49** 47 [*Sov. Phys. - JETP* **22** 33].

[33] Barreda-Argüeso J. A., Aguado F., González J., Valiente R., Nataf L., Sanz-Ortiz M. N., and Rodríguez F. 2016 *J. Phys. Chem. C* **120** 18788.

[34] Loudon R. 1964 *Adv. Phys.* **13** 423.

[35] Bossini D., Kalashnikova A. M., Pisarev R. V., Rasing Th., and Kimel A. V. 2014 *Phys. Rev. B* **89** 60405.

[36] G. Turrell, in Raman Sampling. In Practical Raman Spectroscopy, D.J. Gardiner and P.R. Graves, Eds. (Springer-Verlag, New York, 1989), pp 13–54.

[37] Dhar L., Rogers J. A., and Nelson K. A. 1994 *Chem. Rev.* **94** 157.

[38] Lee C., Ghosez P., and Gonze X. 1994 *Phys. Rev. B* **50** 13379.

[39] Supplimentary Material

[40] Meloche E., Cottam M. G., and Lockwood D. J. 2007 *Phys. Rev. B* **76** 104406.

[41] Hoff J. T., Grünberg P. A., and Koningstein J. A. 1972 *Appl. Phys. Lett.* **20** 358.

[42] Lockwood D. J. and Cottam M. G. 2012 *Low Temp. Phys.* **38** 549.

[43] Meloche E., Cottam M. G., and Lockwood D. J. 2014 *Low Temp. Phys.* **40** 134.

[44] Moch P., Gosso J. P., and Dugautier C., *In Light Scattering in Solids* (Flammarion, Paris, 1971).

[45] Formisano F., Medapalli R., Xiao Y., Ren H., Fullerton E. E., and Kimel A. V. 2020 *J. Magn. Magn. Mater.* **502** 166479.

[46] Nisoli M., De Silvestri S., and Svelto O. 1996 *Appl. Phys. Lett.* **68** 2793.

[47] Arora A. and Ghosh S. 2010 *Rev. Sci. Instrum.* **81** 123102.

[48] Kalashnikova A. M., Kimel A. V., Pisarev R. V., Gridnev V. N., Usachev P. A., Kirilyuk A., and Rasing Th. 2008 *Phys. Rev. B* **78** 104301.

[49] Satoh T., Iida R., Higuchi T., Fiebig M., and Shimura T. 2014 *Nat. Photon.* **9** 25.

[50] Eremenko V. V., Naumenko V. M., Petrov S. V., and Pishko V. V. 1982 *Zh. Eksp. Teor. Fiz.* **82** 813 [*Sov. Phys. - JETP* **55** 481].

[51] Cottam M. G. and Lockwood D. J. 2019 *Low Temp. Phys.* **45** 78.

[52] Thomson R. I., Chatterji T., and Carpenter M. A. 2014 *J. Phys. Condens. Matter* **26** 146001.

[53] Imasaka K., Pisarev R. V., Bezmaternykh L. N., Shimura T., Kalashnikova A. M., and Satoh T. 2018 *Phys. Rev. B* **98** 054303.

[54] Kenney J. F., *Mathematics of Statistics. Part Two*, 2d ed. (D. Van Nostrand Company, Inc., New York, 1947).

[55] Satoh T., Iida R., Higuchi T., Fiebig M., and Shimura T. 2014 *Nat. Photon.* **9** 638.

[56] Porto S. P. S., Fleury P. A., and Damen T. C. 1967 *Phys. Rev.* **154** 522.

[57] Borovik-Romanov A. S., Kreines N. M., Pankov A., and Talalaev M. A. 1973 *J. Exptl. Theor. Phys.* **64** 1762 [*Sov. Phys. - JETP* **37** 890].

[58] Zvezdin A. K. and Kotov V. K., *Modern Magnetooptics and Magnetooptical Materials* (IOP Publishing, Bristol, 1997).

[59] Allen S. J. and Guggenheim H. J. 1971 *Phys. Rev. B* **4** 937.

[60] Gaydamak T. N., Zvyagina G. A., Zhekov K. R., Bilich I. V., Desnenko V. A., Kharchenko N. F., and Fil V. D. 2014 *Low Temp. Phys*. **40** 524.

[61] Sugai S., Shamoto S. I., and Sato M. 1988 *Phys. Rev. B* **38** 6436.

[62] Elliott R. J. and Thorpe M. F. 1969 *J. Phys. C. Solid State Phys.* **2** 1630.

[63] Bossini D., Dal Conte S., Terschanski M., Springholz G., Bonanni A., Deltenre K., Anders F., Uhrig G. S., Cerullo G., and Cinchetti M. 2021 *Phys. Rev. B* **104** 224424

[64] Satoh T., Iida R., Higuchi T., Fujii Y., Koreeda A., Ueda H., Shimura T., Kuroda K., Butrim V. I., and Ivanov B. A. 2017 *Nat. Commun.* **8** 638.

# Laser-induced THz magnetism of antiferromagnetic CoF$_2$

F. Formisano[1*], R. M. Dubrovin[2], R. V. Pisarev[2], A. M. Kalashnikova[2], and A. V. Kimel[1]

[1]Institute for Molecules and Materials, Radboud University, 6571AJ Nijmegen, The Netherlands
[2]Ioffe Institute, 194021 St. Petersburg, Russia
*f.formisano@science.ru.nl

**Supplementary Material**

## 1. Phonon modes in CoF$_2$

CoF$_2$ crystallizes in a tetragonal rutile structure [Fig. 1(a) in the main text], with the space group P4$_2$/mnm (D$^{14}_{4h}$, #136, Z=2) [1]. The unit cell of CoF$_2$ contains six atoms giving rise to 11 optical phonon modes in total with irreducible representations $\varGamma_{\text{optic}} = A_{1g} + A_{2g} + A_{2u} + B_{1g} + B_{2g} + 2B_{1u} + E_g + 3E_u$ [2-4]. All $A$ and $B$ modes are nondegenerate, whereas the $E$ modes are twofold degenerate. The odd $A_{2u}$ and $3E_u$ modes are infrared-active, and within the electric-dipole approximation are not expected to be excited with NIR laser pulses. In contrast, the even $A_{1g}$, $B_{1g}$, $B_{2g}$ and $E_g$ modes are Raman-active. All other phonon modes are silent.

Raman tensors corresponding to the Raman active $A_{1g}$, $B_{1g}$, $B_{2g}$ and $E_g$ modes are

$$\mathcal{R}(A_{1g}) = \begin{pmatrix} R_{aa} & 0 & 0 \\ 0 & R_{aa} & 0 \\ 0 & 0 & R_{cc} \end{pmatrix};$$

$$\mathcal{R}(E_g) = \begin{pmatrix} 0 & 0 & 0 \\ 0 & 0 & R_{bc} \\ 0 & R_{bc} & 0 \end{pmatrix}, \mathcal{R}(E_g) = \begin{pmatrix} 0 & 0 & R_{ac} \\ 0 & 0 & 0 \\ R_{bc} & 0 & 0 \end{pmatrix}; \quad (S1)$$

$$\mathcal{R}(B_{1g}) = \begin{pmatrix} R_{aa} & 0 & 0 \\ 0 & -R_{aa} & 0 \\ 0 & 0 & 0 \end{pmatrix}, \mathcal{R}(B_{2g}) = \begin{pmatrix} 0 & R_{ab} & 0 \\ R_{ab} & 0 & 0 \\ 0 & 0 & 0 \end{pmatrix},$$

with nonzero elements showing what combination of incident and scattered light enable observation of these modes in the Raman scattering experiments. As Raman active modes, these modes can be driven by NIR laser pulses via the ISRS mechanism [5-7]. In particular, pump pulse propagating along the $b$-axis and polarized in the $ac$-plane can excite all of these modes except of $B_{2g}$. In experiments, we have indeed observed excitation of $A_{1g}$, $B_{1g}$, and $E_g$ modes (Fig. 2(b) in the main text).

## 2. Probe polarization rotation modulation by the coherent $E_g$ phonon

*2.1. Linear birefringence induced by the coherent $E_g$ phonon*

Static dielectric tensor $\varepsilon_{ij}$ of $CoF_2$ in the paramagnetic phase is symmetric and has only diagonal components $\varepsilon_{aa}^{(s)\prime} = \varepsilon_{bb}^{(s)\prime} \neq \varepsilon_{cc}^{(s)\prime}$. In the spectral range where absorption can be neglected, these components are real and give rise to the crystallographic birefringence for the light propagating perpendicular to the $c$ axis. In our experiment, pump and probe pulses propagate close to the $b$ axis. In the antiferromagnetic phase, there are additional contributions to the diagonal components $\varepsilon_{aa}^{(s)} = \varepsilon_{bb}^{(s)} = \varepsilon_{aa}^{(s)\prime} + c_1 L^2$ and $\varepsilon_{cc}^{(s)} = \varepsilon_{cc}^{(s)\prime} + c_2 L^2$ [8], where $L$ is the antiferromagnetic vector. The eigen waves propagating along the $b$ axis are two waves linearly polarized along the $a$ and $c$ axes, and the corresponding refractive indices are $n_a = n_b = \sqrt{\varepsilon_{aa}^{(s)}}$ and $n_c = \sqrt{\varepsilon_{cc}^{(s)}}$.

Once the coherent phonon mode of the $E_g$ symmetry with the normal coordinate $Q(t)$ is excited by the pump pulse, the dielectric permittivity tensor $\varepsilon_{ij}^{(s)}$ is expected to acquire the off-diagonal components $\varepsilon_{ac}^{(s)}(t) = \varepsilon_{ca}^{(s)}(t) = R_{ac} Q(t)$ [see Eq. (4) in the main text]. Then the eigen waves for the medium described by Eq. (4) are still the two linearly polarized waves with electric fields making the angle $\alpha(t)$ with the crystallographic $a$ and $c$ axes:

$$\tan\alpha(t) = \frac{2 R_{ac} Q(t)}{(\varepsilon_{aa}^{(s)}-\varepsilon_{cc}^{(s)})-(\varepsilon_{aa}^{(s)}-\varepsilon_{cc}^{(s)})\sqrt{1+\left(\frac{2R_{ac}Q(t)}{\varepsilon_{aa}^{(s)}-\varepsilon_{cc}^{(s)}}\right)^2}}. \tag{S2}$$

Corresponding refraction indices are

$$n_{1(2)}^2(t) = \frac{\varepsilon_{aa}^{(s)}+\varepsilon_{cc}^{(s)}}{2} \pm \frac{\varepsilon_{aa}^{(s)}-\varepsilon_{cc}^{(s)}}{2}\sqrt{1+\left(\frac{2R_{ac}Q(t)}{\varepsilon_{aa}^{(s)}-\varepsilon_{cc}^{(s)}}\right)^2}. \tag{S3}$$

Assuming the modulation of the dielectric permittivity by the coherent optical phonon to be considerably weaker than the static birefringence, i.e. $|R_{ac} Q(t)| \ll |\varepsilon_{cc}^{(s)} - \varepsilon_{aa}^{(s)}|$, one gets the approximate expressions for $\alpha(t)$ and the refractive indices $n_{1(2)}(t)$ for the eigen waves [9]

$$\alpha(t) \approx \frac{2R_{ac}Q(t)}{\varepsilon_{aa}^{(s)}-\varepsilon_{cc}^{(s)}}, \tag{S4a}$$

$$n_{1(2)}(t) \approx \sqrt{\varepsilon_{aa(cc)}^{(s)}} \pm \frac{2R_{ac}^2 Q(t)^2}{\sqrt{\varepsilon_{aa(cc)}^{(s)}}\left(\varepsilon_{aa}^{(s)}-\varepsilon_{cc}^{(s)}\right)^2}. \tag{S4b}$$

Equations (S4) show that the presence of the coherent $E_g$ phonon mode results in periodic rotational oscillation of the basis formed by the two eigen polarizations at the phonon frequency. The corresponding refractive indices are, in turn, modulated at the doubled frequency. Thus, the main effect of the excited coherent $E_g$ phonon mode on the probe polarization at the phonon frequency $f_0$ comes from the rotation of the basis formed by eigen polarizations [9]. Therefore, below we take $n_{1(2)} \approx n_{a(c)} = \sqrt{\varepsilon_{aa(cc)}^{(s)}}$, and $\Delta n = \sqrt{\varepsilon_{aa}^{(s)}} - \sqrt{\varepsilon_{aa}^{(s)}}$. Note, that for the case of the medium with negligible static birefringence ($\varepsilon_{aa}^{(s)} = \varepsilon_{bb}^{(s)} = \varepsilon_{cc}^{(s)} = \varepsilon_0^{(s)}$), Eqs. (S2,S3) yield that the eigen waves in the presence of the $E_g$ phonon are the waves polarized at $\alpha(t) = 45$ deg with the corresponding refraction indices $n_{1(2)}(t) = \sqrt{\varepsilon_0^{(s)} \pm R_{ac}Q(t)}$.

To describe the propagation of the probe pulse through a material with such properties we write down the Jones matrix for the slice of the crystal of thickness $\delta z$ at the distance $z$ along the $b$ axis

$$\mathbf{T}(\delta z) = \begin{bmatrix} \exp\left[-i\frac{2\pi\Delta n \delta z}{\lambda}\right] & 0 \\ 0 & 1 \end{bmatrix}. \tag{S5}$$

Equation (S5) is written in the coordinate frame linked to the transient axes rotated from the $a$ and $c$ crystallographic axes by $\alpha(t)$. In order to find the modulation of $\theta(\delta z)$ when $\alpha(t) \neq 0$, we note that $\alpha(t)$ can be replaced by small rotation of the incoming probe polarization $\delta\theta = -\alpha(t)$. Then the corresponding change of the probe polarization after the probe propagated through the sample can be found as $\Delta\theta(\delta z) = [\partial[\theta(\delta z) - \theta]/\partial\theta] \cdot \delta\theta$.

Expression of $\theta(\delta z)$ is derived by finding the Jones vector

$$\begin{bmatrix} E_a(\delta z) \\ E_c(\delta z) \end{bmatrix} = \mathbf{T}(\delta z) \cdot \begin{bmatrix} E_0 \cos\theta \\ E_0 \sin\theta \end{bmatrix} \tag{S6}$$

for the light after traveling the distance $\delta z$ through a medium and the corresponding polarization rotation $\theta(\delta z)$ and ellipticity $\psi(\delta z)$ [10]

$$\tan 2\theta(\delta z) = \frac{2\mathrm{Re}\chi(\delta z)}{1-|\chi(\delta z)|^2}, \tag{S7a}$$

$$\sin 2\psi(\delta z) = \frac{2\mathrm{Im}\chi(t,\delta z)}{1+|\chi(\delta z)|^2}, \tag{S7b}$$

$$\chi(\delta z) = \frac{E_c(\delta z)}{E_a(\delta z)}. \tag{S7c}$$

The probe polarization rotation $\theta(\delta z)$ with respect to the incoming polarization, $\theta$, is then expressed as

$$\theta(\delta z) = \frac{1}{2}\mathrm{arctg}\left[\cos\frac{2\pi\Delta n \delta z}{\lambda}\tan 2\theta\right] - \theta. \tag{S8}$$

Note, that Eq. (S8) gives a trivial solution $\theta(\delta z) = 0$ for an isotropic medium ($\Delta n=0$). The change of the probe polarization after the probe propagated the distance $\delta z$ is:

$$\Delta\theta(t,\delta z) = \alpha(t)\left[1 - \cos\frac{2\pi\Delta n\delta z}{\lambda}\left[\cos^2 2\theta + \left(\cos\frac{2\pi\Delta n\delta z}{\lambda}\sin 2\theta\right)^2\right]^{-1}\right]. \qquad (S9)$$

Equation (S9) yields that, for instance, $\Delta\theta(t,\delta z) = \alpha(t)\left[1 - \cos\frac{2\pi\Delta n\delta z}{\lambda}\right]$ at $\theta=0$ deg, and $\Delta\theta(t,\delta z) = \alpha(t)\left[1 - \sec\frac{2\pi\Delta n\delta z}{\lambda}\right]$ at $\theta=45$ deg. Thus, the transient probe polarization rotation is larger when the incoming probe polarization is at $\theta=45$ deg to the crystallographic axes. Indeed, as follows from Eq. (S9), in isotropic medium $\Delta\theta(t,\delta z)=0$. $\Delta\theta(t,\delta z)$ occurs only because the static crystallographic birefringence converts the transient ellipticity induced by the coherent phonon to the rotation. Such conversion is the most efficient when the incoming probe polarization at $\theta=45$ deg, and the static birefringence is the largest. Note, that the birefringence $\Delta n$ should exceed $\approx 5\cdot 10^{-4}$ to provide a noticeable increase of the transient rotation $\Delta\theta(t,d)$ at $\theta=45$ deg. The birefringence in $CoF_2$ meets with this condition.

It is important to take into account that in birefringent media the light polarization changes upon propagation through the sample. In our experiments pump pulse is initially polarized at $\phi=45$ deg to the $a$ axis, and, thus, becomes elliptically polarized with ellipticity $\psi_{\text{pump}}(\delta z) = \pi\Delta n\delta z/\lambda$. This results in different efficiency of the excitation of the coherent phonon along the sample depth, and, thus, in $Q(t)$ and $\alpha(t)$ being dependent on $\delta z$. This effect can be neglected, and $Q(t)$ and $\alpha(t)$ can be considered as being independent of $\delta z$ only if the ellipticity acquired by the pump pulses at $\delta z = d$ is not large, e.g. $\Delta nd < \lambda/8$, or $\Delta n < 1.8\cdot 10^{-4}$ for $d=500$ μm and $\lambda=800$ nm.

In $CoF_2$ the latter condition is not satisfied, and inhomogeneous excitation of the coherent phonon should be taken into account as $\alpha(t,z) \sim Q(t,z) \sim \cos\psi_{\text{pump}}(z) = \cos(\pi\Delta nz/\lambda)$, and the transient probe polarization rotation should be obtained numerically. In order to do so, the sample is split into thin slices $\delta z$, $\Delta n\delta z \ll \lambda/8$, within which $\alpha(t,z)$ can be considered as constant. The probe polarization after each slice is found as

$$\begin{bmatrix} E_a(t,z+\delta z) \\ E_c(t,z+\delta z) \end{bmatrix} = \mathbf{T}'(t,z)\cdot\begin{bmatrix} E_a(t,z) \\ E_c(t,z) \end{bmatrix}, \qquad (S10)$$

where

$$\mathbf{T}'(t,z) = \text{rot}(-\alpha(t,z))\cdot\mathbf{T}(\delta z)\cdot\text{rot}(+\alpha(t,z)) \qquad (S11)$$

$$= \begin{bmatrix} \exp\left[-i\frac{2\pi\Delta n\delta z}{\lambda}\right]\cos^2\alpha(t,z) + \sin^2\alpha(t,z) & \frac{1}{2}\left(\exp\left[-i\frac{2\pi\Delta n\delta z}{\lambda}\right] - 1\right)\sin 2\alpha(t,z) \\ \frac{1}{2}\left(\exp\left[-i\frac{2\pi\Delta n\delta z}{\lambda}\right] - 1\right)\sin 2\alpha(t,z) & \exp\left[-i\frac{2\pi\Delta n\delta z}{\lambda}\right]\sin^2\alpha(t,z) + \cos^2\alpha(t,z) \end{bmatrix}$$

$$\approx \begin{bmatrix} \exp\left[-i\frac{2\pi\Delta n\delta z}{\lambda}\right] & \left(\exp\left[-i\frac{2\pi\Delta n\delta z}{\lambda}\right]-1\right)\alpha(t,z) \\ \left(\exp\left[-i\frac{2\pi\Delta n\delta z}{\lambda}\right]-1\right)\alpha(t,z) & 1 \end{bmatrix},$$

is written in the crystallographic axes, and $\text{rot}(\pm\alpha(t,z))$ is the matrix describing rotation around the $b$ axis by the angle $\pm\alpha(t,z)$. In the last line in Eq. (S11), the terms quadratic in small deviation $\alpha(t,z)$ are omitted. Using Eqs. (S10, S11) and performing iterative calculations for $z=0\ldots d$, one finds the Jones vector $\mathbf{E}(t,d)$ for the probe after propagation through the sample. Corresponding $\theta(t,d)$ and $\psi(t,d)$ and $\Delta\theta(t,d)=\theta(t,d)-\theta$ are then found using Eqs. (S7).

## 2.2. Circular birefringence induced by the coherent $E_g$ phonon

Rotation of the polarization of light propagating along the magnetization in the medium with linear crystallographic birefringence has been analyzed in detail in Refs. [11,12]. In the case of a medium described by the dielectric permittivity tensor with symmetric and antisymmetric parts [Eqs. (4,6) in the main text], the polarization of light after propagation through the sample of thickness $d$ can be found from [11]

$$\begin{bmatrix} E_a(t,d) \\ E_c(t,d) \end{bmatrix} = \mathbf{T}''(t,z) \cdot \begin{bmatrix} E_a(t,0) \\ E_c(t,0) \end{bmatrix}; \tag{S12a}$$

$$\mathbf{T}''(t,d) \approx \begin{bmatrix} \exp\left[-i\frac{2\pi\Delta nd}{\lambda}\right] & -\frac{\sqrt{2}PQ(t)}{2\sqrt{\varepsilon_{cc}^{(s)}-\varepsilon_{aa}^{(s)}}}\sin\frac{\pi\Delta nd}{\lambda} \\ \frac{\sqrt{2}PQ(t)}{2\sqrt{\varepsilon_{cc}^{(s)}-\varepsilon_{aa}^{(s)}}}\sin\frac{\pi\Delta nd}{\lambda} & 1 \end{bmatrix}, \tag{S12b}$$

where it was assumed that $PQ(t) \ll \varepsilon_{cc}^{(s)}-\varepsilon_{aa}^{(s)}$. Numerical simulations using Eqs. (S7) and (S12) show that for small values of $PQ(t)$, i.e. small specific Faraday rotation emerging due to the coherent phonon, corresponding rotation of the probe polarization would be larger at $\theta$=45, 135 deg as compared to $\theta$=90, 135 deg. In particular, for $\Delta n \approx 3\cdot 10^{-4}$ and $PQ(t)\approx 5\cdot 10^{-5}$ one gets that the rotation experienced by the probe pulse initially polarized at $\theta$=45 deg is 2 times larger compared to the case with probe pulse polarized along one of the axes [see inset in Fig. S1(b)].

Figure S1 shows the experimental dependence of the signed amplitude $\Delta\theta_0$ of the 7.45 THz oscillations mode on the temperature at different incoming probe polarizations. The temperature-dependent contribution is the most apparent and possesses the same, negative, sign when the incoming polarization is close to $\theta=45$ to $\theta=135$ deg, [see inset in Fig. S1(b)]. To highlight the qualitative agreement with the results of the analysis using Eq. (S12), we have also plotted in Fig. S1(a) the curve obtained at $\theta=45$ deg multiplied by the factor of 0.5. This curve indicates the change of the signal expected at $\theta$ close to 90 deg, according to the simulations described above. One can see that such a change cannot be resolved in the experiment due to an insufficient signal-to-noise ratio. If the effect of birefringence can be neglected, the rotation of the probe polarization can be found as in the case of isotropic medium [13] with a refraction index of $0.5\sqrt{\varepsilon_{cc}^{(s)} + \varepsilon_{aa}^{(s)}}$ [see Eq. (7) in the main text].

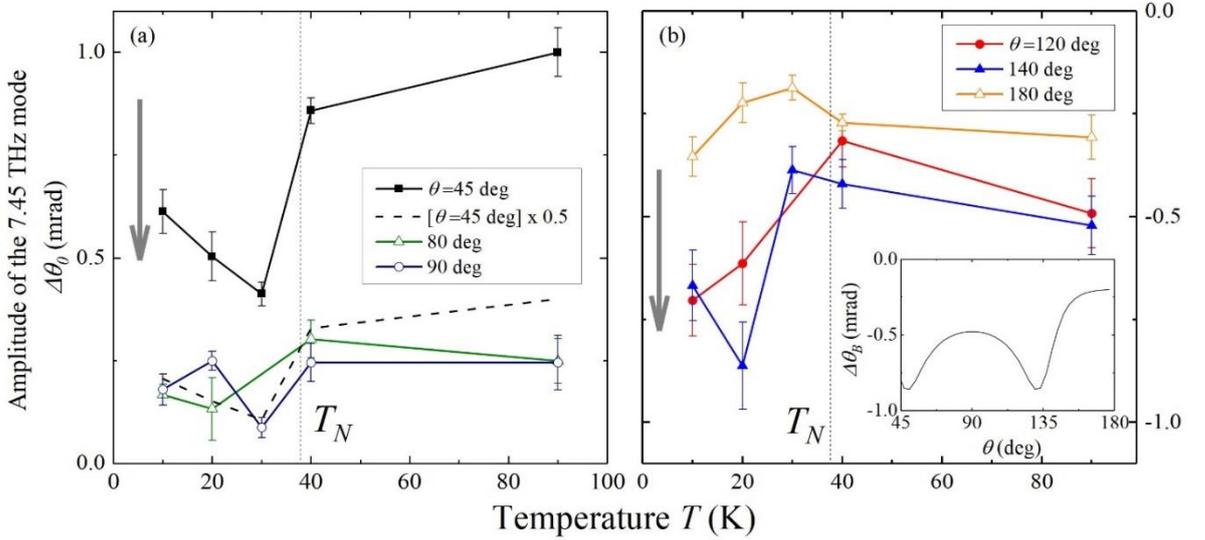

**Figure S1**. Temperature dependencies of the amplitude of the oscillations $\Delta\theta_0$ at the frequency of 7.45 THz seen in the rotation of the probe polarization. The measurements were performed at different incoming probe polarizations $\theta$. Left (a) and right (b) panels show the data obtained at $\theta=45$, 80 and 90 deg and 120, 140 and 180 deg, respectively. Grey vertical arrows highlight changes of $\Delta\theta_0$ when $\theta$ is close to 45 or 135 deg. Inset in panel (b) shows calculated dependence of the transient rotation $\Delta\theta_B$ on the incoming probe polarizations $\theta$ in the birefringent medium (see text for details). Dashed line in panel (a) shows the curve obtained at $\theta=45$ deg multiplied by the factor of 0.5.

## 3. Phonon-induced transverse piezomagnetic effect

Following Ref. [14] we write down the expression for the magnetization-dependent part of the free energy of $CoF_2$ in the presence of shear strains $\Lambda_{ac}^i$ and $\Lambda_{bc}^i$, where $i=1,2$ for two sublattices:

$$F = \frac{1}{2}A_1[(m_{1a} - m_{2a})^2 + (m_{1b} - m_{2b})^2] + \frac{1}{2}A_2(\mathbf{m_1} + \mathbf{m_2})^2 + \frac{1}{2}d_1(m_{1c} + m_{2c})^2$$
$$+ d_2[(m_{1a} + m_{2a})(m_{1b} - m_{2b}) + (m_{1b} + m_{2b})(m_{1a} - m_{2a})]$$
$$+ \lambda_1(m_{1a}\Lambda_{bc}^1 + m_{2a}\Lambda_{bc}^2 + m_{1b}\Lambda_{ac}^1 + m_{1b}\Lambda_{ac}^2)(m_{1c} - m_{2c})$$
$$+ \eta_1(m_{1a}\Lambda_{ac}^1 - m_{2a}\Lambda_{ac}^2 + m_{1b}\Lambda_{bc}^1 - m_{2b}\Lambda_{bc}^2)(m_{1c} - m_{2c}). \tag{S13}$$

Static shear strains have the same sign for the $Co^{2+}$ ions at both sublattices ($\Lambda_{ac}^1 = \Lambda_{ac}^2 = \Lambda_c$, $\Lambda_{bc}^1 = \Lambda_{bc}^2 = \Lambda_{bc}$) and result in the emergence of the net magnetic moment

$$\delta\mathbf{m} = \delta\mathbf{m_1} + \delta\mathbf{m_2} = \frac{d_2\eta_1 - A_1\lambda_1}{A_1A_2 - d_2^2}\Lambda_{bc}\mathbf{a} + \frac{d_2\eta_1 - A_1\lambda_1}{A_1A_2 - d_2^2}\Lambda_{ac}\mathbf{b}. \tag{S14}$$

In contrast to the static shear strains, time-dependent local shear distortions associated with the $E_g$ phonon mode for the $Co^{2+}$ ions belonging to different sublattices have the same magnitudes but are related to each other through the 90° rotation around the c axis $\Lambda_{ac}^1(t) = \Lambda_{ac}^2(t) = \Lambda_{ac}(t) \sim Q(t)$ and $\Lambda_{bc}^1(t) = -\Lambda_{bc}^2(t) = \Lambda_{ac}(t) \sim Q(t)$ [Fig. 1(a) in the main text]. Taking this into account we obtain for the shear-distortions dependent part of Eq. (S13):

$$F = \lambda_1\big(m_{1a}\Lambda_{bc}(t) - m_{2a}\Lambda_{bc}(t) + m_{1b}\Lambda_{ac}(t) + m_{1b}\Lambda_{ac}(t)\big)(m_{1c} - m_{2c})$$
$$+ \eta_1\big(m_{1a}\Lambda_{ac}(t) - m_{2a}\Lambda_{ac}(t) + m_{1b}\Lambda_{bc}(t) + m_{1b}\Lambda_{bc}(t)\big)(m_{1c} - m_{2c}), \tag{S15}$$

that reduces to Eq. (8) in the main text after introducing net magnetization **M** and antiferromagnetic vector **L**.

**References**

[1] Stout J. W. and Reed S. A. 1954 *J. Am. Chem. Soc.* **76** 5279.

[2] Loudon R. 1964 *Adv. Phys.* **13** 423.

[3] Porto S. P. S., Fleury P. A., and Damen T. C. 1967 *Phys. Rev.* **154** 522.

[4] Lee C., Ghosez P., and Gonze X. 1994 *Phys. Rev. B* **50** 13379.

[5] Yan Y. X., Gamble E. B., and Nelson K. A. 1985 *J. Chem. Phys.* **83** 5391.

[6] Merlin R. 1997 *Solid State Commun.* **102** 207.

[7] Dhar L., Rogers J. A., and Nelson K. A. 1994 *Chem. Rev.* **94** 157.

[8] Borovik-Romanov A. S., Kreines N. M., Pankov A., and Talalaev M. A. 1973 *J. Exptl. Theor. Phys.* **64** 1762 [*Sov. Phys. - JETP* **37** 890].

[9] Imasaka K., Pisarev R. V., Bezmaternykh L. N., Shimura T., Kalashnikova A. M., and Satoh T. 2018 *Phys. Rev. B* **98** 054303.

[10] Azzam R. M. A. and Bashara N. M., *Ellipsometry and Polarized Light* (North-Holland, 1988).

[11] Tabor W. J. and Chen F. S. 1969 *J. Appl. Phys*. **40** 2760.

[12] Kahn F. J., Pershan P. S., and Remeika J. P. 1969 *Phys. Rev.* **186** 891.

[13] Zvezdin A. K. and Kotov V. K., *Modern Magnetooptics and Magnetooptical Materials* (IOP Publishing, Bristol, 1997).

[14] Borovik-Romanov A. S. 1960 *J. Exptl. Theor. Phys*. **38** 1088 [*Sov. Phys. - JETP* **11** 786].